# Absorption cross-section spectroscopy of single strong coupling system between plasmon and molecular exciton resonance using single silver nanoparticle dimer generating surface enhanced resonant Raman scattering

Tamitake Itoh[1]*, Yuko S. Yamamoto[2], Takayuki Okamoto[3]

[1] Nano-Bioanalysis Research Group, Health Research Institute, National Institute of Advanced Industrial Science and Technology (AIST), Takamatsu, Kagawa 761-0395, Japan

[2] School of Materials Science, Japan Advanced Institute of Science and Technology (JAIST), Nomi, Ishikawa 923-1292, Japan

[3] Advanced Device Laboratory, RIKEN, Hirosawa, Wako, Saitama 351-0198, Japan

*Corresponding author: tamitake-itou@aist.go.jp

**ABSTRACT**

This study investigated spectral changes in the absorption cross-sections of single strong coupling systems composed of single silver nanoparticle dimers and a few dye molecules during the quenching of surface-enhanced resonant Raman scattering (SERRS). The absorption cross-section was obtained by subtracting the scattering cross-section from an extinction cross-section. The spectral changes in these cross-sections were evaluated using a classical hybridization model composed of a plasmon and a molecular exciton including a molecular multi-level property. The changes in the scattering and extinction cross-sections exhibit blue-shifts in their peak energy and increased peak intensities, respectively, during SERRS quenching. These properties are effectively reproduced in the model by decreasing the coupling energy. In particular, the peaks in the scattering and extinction cross-sections appear as peaks or dips in the absorption cross-sections depending on the degree of scattering loss, which reflects the dimer sizes. These results are useful for optimizing photophysical and photochemical effects mediated by the electronic excited states of strong coupling systems.

## I. INTRODUCTION

Following the discovery of single-molecule (SM) surface-enhanced resonant Raman scattering (SERRS) [1,2], it has become possible to investigate the relationships between SERRS and plasmon resonance directly, because one can identify the plasmon "resonance" inducing the Raman "enhancement" [3-6]. Such investigations have revealed that plasmonic nanoparticle (NP) dimers or NPs on plasmonic substrates generating SERRS of single or near-single dye molecules form strong coupling systems, in which the electromagnetic (EM) coupling rates between plasmon resonance and molecular exciton resonance are greater than the dephasing rates of both resonances [7-12]. The reported EM coupling energies reach several hundred meV, which corresponds to almost one cubic nanometer of the mode volume of a confined field by plasmon resonance at junctions called "hotspots" [13]. In the cavity quantum electrodynamics (cavity QED) model, the EM coupling hybridizes plasmon resonance with molecular exciton resonance, and the hybridized resonance changes the energies by half of $2\hbar g$, where $g$ is the EM coupling rate, from the original resonance energies under the rigorous resonant condition shown in Fig. 1(a) [13,14]. The changes in resonance energies appear as spectral splitting by $2\hbar g$, as shown in Fig. 1(b) [15]. For each molecule involved in the EM coupling, the change in resonance energy corresponds to $\frac{2\hbar g}{\sqrt{N}}$, where $N$ is the number of dye molecules involved in the EM coupling [15]. Thus, this change is negligible for large values of $N$, indicating that the resonance energy of each molecule does not change [15]. However, in the SM SERRS case of $N \sim 1$, $\frac{2\hbar g}{\sqrt{N}}$ may not be negligible, suggesting that the molecule will exhibit resonant like photophysical or photochemical responses, even under conventional non-resonant excitation conditions, because of the change in resonant energy. Indeed, there are several reports of unique pumping effects and photochemical effects for plasmonic systems [16,17]. Therefore, it is important to investigate the optical absorption of strong coupling systems to elucidate these effects [18].

The optical absorption of collective strong coupling systems has been intensively studied for plasmonic NPs on plasmonic substrates with a rather high dye molecule concentration condition, i.e., $N \gg 1$ [15,19]. In the case of $N \sim 1$, as in near-SM SERRS systems, any fluctuation in $N$ induces large variations in the EM coupling energies, because $\hbar g$ is proportional to $\sqrt{N}$. Furthermore, the detailed changes in the structures of hotspots and the molecular positions in these hotspots may result in large variations in EM coupling energies [12]. Plasmonic NP systems also exhibit considerable system-by-system variations in their resonance energies and line widths, reflecting their sizes and shapes [5]. Therefore, by excluding such variations, the development of optical absorption spectroscopy for single strong coupling systems is an important step in correctly evaluating the relationships between these unique phenomena and strong coupling. The other intrinsic importance of investigating single strong coupling systems is the identification of the plasmon "resonance" that induces the strong coupling [9,12]. In the case of SERRS, the relationships between plasmon resonance and SERRS have been confirmed using single silver NP dimers [5,20-23]. Furthermore, direct measurement of quantum efficiency of resonant light scattering of SERRS hotspots will be possible by comparing between scattering with absorption cross-section [13,24].

In this study, we develop a spectroscopic method for the absorption cross-sections $\sigma_{abs}(\omega)$ of single strong coupling systems composed of a silver NP dimer and a few dye molecules. We then investigate $\sigma_{abs}(\omega)$ during the SERRS quenching process, where $\omega$ is the angular frequency of incident light. The spectra of the scattering and extinction intensities of silver NP dimers are converted into cross-sections using reference spectra for gold NPs, where the scattering cross-sections $\sigma_{sca}(\omega)$ and extinction cross-sections $\sigma_{ext}(\omega)$ have been quantitatively reproduced by Mie theory. We then obtain $\sigma_{abs}(\omega)$ for the dimer by subtracting $\sigma_{sca}(\omega)$ from $\sigma_{ext}(\omega)$. The spectral changes in $\sigma_{sca}(\omega)$, $\sigma_{ext}(\omega)$, and $\sigma_{abs}(\omega)$ are evaluated using a classical hybridization model, which gives essentially identical results to a QED model of vacuum EM fields under a quasi-static approximation. The model effectively reproduces the experimental spectral changes during SERRS quenching by reducing the EM coupling energies. The difference in the intensity changes in $\sigma_{sca}(\omega)$ and $\sigma_{ext}(\omega)$ are explained through their size parameter dependence. The appearance of peaks and dips in $\sigma_{abs}(\omega)$ at the peak positions of $\sigma_{sca}(\omega)$ and $\sigma_{ext}(\omega)$ is clarified as the dimer-size-dependent scattering loss. Using the model, the considerable variations in the spectral changes for different dimers are comprehensively explained as a function of the peak energies of plasmon resonance, coupling energies, and dimer sizes.

## II. EXPERIMENT

Colloidal gold NPs (mean diameters 60, 80, and 100 nm, standard deviation 8 nm, EMGC40, Funakoshi, Japan) were used to investigate the conversion from scattering and extinction intensities to their cross-sections. A calculation algorism based on Mie theory [25] with a dielectric function of gold [26] quantitatively reproduces the experimental optical properties of gold NPs [27]. Colloidal silver NPs (mean diameter 40 nm, $1.10 \times 10^{-10}$ M) were prepared for a SERRS experiment using the Lee and Meisel method [28]. The colloidal silver NP dispersion was added to the same amount of R6G aqueous solutions ($1.28 \times 10^{-8}$ M) with NaCl (8 mM) and left for 30 min for SERRS activation. The sample solution (50 μL) was then dropped onto a slide glass plate, and the drop was sandwiched by a cover glass plate to immobilize the SERRS active colloidal silver NPs on the plate. This sample plate was then set on the stage of an inverted optical microscope (IX-71, Olympus, Tokyo).

The scattering and extinction spectra of single gold NPs and silver NPs were measured by illuminating white light from a 50-W halogen lamp through a dark-field condenser (numerical aperture (N.A.) 0.92). In measuring the scattering (extinction) spectra, the N.A. of the objective lens (LCPlanFL 100×, Olympus, Tokyo) was set to be 0.6 smaller (1.3 larger) than the N.A. of the dark-field condenser to realize dark- (bright-) field illumination. Figures 1(c)–1(e) show a dark-field, bright-field, and SERRS image of the same sample area of a glass plate surface, respectively. Silver NPs appear as colorful bright and dark spots reflecting the sizes, shapes, and aggregation of the NPs. The scattering and extinction spectra of single NPs were measured by selecting one spot in the image using a pinhole in front of a polychromator equipped with a thermoelectrically cooled charge-coupled device (CCD) (DV 437-OE-MCI, Andor, Japan) assembly. The SERRS image was measured by focusing an excitation green laser beam (2.33 eV (532 nm), 3.5 W/cm$^2$) from a CW Nd3+:YAG laser (DPSS 532, Coherent, Tokyo) on the sample plate through an objective lens (5×, N.A. 0.15, Olympus, Tokyo) [9]. The SERRS active silver NPs are always aggregated, rather than appearing as isolated NPs. If we select SERRS active silver NP aggregates showing dipolar plasmon resonance with maxima of 1.7–2.1 eV, such aggregates are always dimers [5,6]. Figures 1(f) and 1(g) show the plasmon resonance spectra of $\sigma_{sca}(\omega)$,

SERRS spectra, and SEM images of two typical SERRS active silver NP dimers.

Note that the final concentrations of the R6G solution ($6.34 \times 10^{-9}$ M) and NP dispersion ($5.5 \times 10^{-11}$ M) are close to the SM SERRS condition examined by a two-analyte and isotope technique [29,30]. Thus, we believe that the number of dye molecules involved in the present strong coupling may be quite small ($< 6$) [12]. Indeed, we frequently observed signal fluctuation and blinking behaviors in SERRS, which is indirect evidence of SM observations [6].

## III. RESULTS AND DISCUSSION

We compared the experimental plasmon resonance spectra for the scattering and extinction of gold NPs with those calculated using Mie theory to convert the scattering and extinction intensities into their cross-sections. Mie theory quantitatively reproduces the plasmon resonance properties of spherical-like gold NPs using a suitable dielectric function of gold and uniform environmental conditions [27,31]. Thus, gold NPs on a glass plate are covered with an index matching oil with a refractive index $n_{\text{ref}} = 1.518$, which is the same as that of the glass plate, to ensure that the refractive index of the surrounding media is uniform. Figures 2(a)–2(c) are the calculated spectra of $\sigma_{\text{sca}}(\omega)$, $\sigma_{\text{ext}}(\omega)$, and $\sigma_{\text{abs}}(\omega)$ for gold NPs with diameters of 30, 70, and 100 nm, respectively. The following relationship between the cross-sections and diameters can be observed:

$\sigma_{ext}(\omega) \sim \sigma_{abs}(\omega) \gg \sigma_{sca}(\omega)$ for the 30-nm NPs, $\sigma_{ext}(\omega)/2 \sim \sigma_{abs}(\omega) \sim \sigma_{abs}(\omega)$ for the 70-nm NPs, and $\sigma_{abs}(\omega) \ll \sigma_{abs}(\omega) \sim \sigma_{ext}(\omega)$ for the 100-nm NPs. The size dependence of the relationship among $\sigma_{abs}(\omega)$, $\sigma_{sca}(\omega)$, and $\sigma_{ext}(\omega)$ can be explained by the following quasi-static approximation, which assumes that individual NPs are too small to cause any retardation effect: $\sigma_{ext} \propto x$, $\sigma_{sca} \propto x^4$, and $\sigma_{ext}(\omega) = \sigma_{abs}(\omega) + \sigma_{sca}(\omega)$, where $x$ is a size parameter expressed as $x = \dfrac{\omega n_{ref} d}{2c}$ [25]. Here, $d$ is the NP diameter and $c$ is the velocity of light. In other words, $\sigma_{ext} \propto x$ and $\sigma_{sca} \propto x^4$ indicate that the scattering loss in $\sigma_{ext}(\omega)$ is negligible for small NPs, but is dominant for large NPs.

The size dependence of $\sigma_{sca}(\omega)$, $\sigma_{ext}(\omega)$, and $\sigma_{abs}(\omega)$s was experimentally examined using gold NPs of three diameters. Figure 2(d) shows the relationships between the plasmon resonance (dipole mode) peak energies and their cross-sections for gold NPs with diameters of 60, 80, and 100 nm and the relationships calculated by Mie theory using gold NPs with diameters from 20–100 nm. We selected the dipole plasmon mode for the analysis, because the analysis will be applied to SERRS active silver NP dimers exhibiting dipolar plasmon resonance [7,10]. The redshifts observed in the peak experimental energies with increasing intensity are quantitatively reproduced in the calculations by increasing the NP diameters. The tendencies for both $\sigma_{ext} \propto x$ and $\sigma_{ext} \propto x^4$ are observed in the experiments, and the saturation and decrease in $\sigma_{abs}(\omega)$ can also be seen. This quantitative consistency between the experiments and calculations indicates that we can convert experimental spectral intensities into their cross-sections based on the Mie theoretical calculations.

Figures 2(e)–2(g) show the experimental and calculated scattering, extinction, and absorption spectra for gold NPs with diameters of 60, 80, and 100 nm. The experimental spectra for $\sigma_{abs}(\omega)$ are obtained by subtracting $\sigma_{sca}(\omega)$ from $\sigma_{ext}(\omega)$ [19,25]. The experimental results are reproduced reasonably well by the calculations except in the higher-energy region (>2.5 eV) for the 100-nm diameter NPs. This failure is due to incomplete detection of the extinction spectra of higher-order plasmon modes such as quadrupoles, resulting in the underestimation of $\sigma_{abs}(\omega)$ in the higher-energy region. This incomplete detection is in turn caused by the configuration of our spectroscopic system, because conventional and photothermal absorption spectroscopy can correctly measure the extinction spectra of the higher-energy region [19,27,32]. Given these comparisons, we consider that all experimental spectra measured by the current spectroscopic system can be converted into cross-section spectra in the same manner within the spectral region where the dipole plasmon mode is dominant. Note that the present determination method of cross-sections is not applicable for large NPs because of the overestimation of $\sigma_{sca}(\omega)$. Indeed, we use only forward scattering intensities to derive $\sigma_{sca}(\omega)$ without considering the fact that forward-scattering intensity becomes larger than back-scattering intensity with increasing NP sizes, even the calculated $\sigma_{sca}(\omega)$ include such fact [25].

We applied the above conversion method to the plasmon resonance spectra from 14 silver NP dimers before and after SERRS quenching. Figures 3(a)–3(c) show the representative experimental plasmon resonance spectra of $\sigma_{abs}(\omega)$, $\sigma_{sca}(\omega)$, $\sigma_{ext}(\omega)$ and SERRS spectra for three silver NP dimers before and after the loss of SERRS activity, respectively. Figures 3(a1)–3(c1) and 3(a2)–3(c2) shows the cross-section spectra before and after the loss of SERRS activity, respectively. The current dimer experiments can be classified into three categories: $\sigma_{abs}(\omega) > \sigma_{sca}(\omega)$ for Fig. 3(a), $\sigma_{abs}(\omega) \sim \sigma_{sca}(\omega)$ for Fig. 3(b), and $\sigma_{abs}(\omega) < \sigma_{sca}(\omega)$ for Fig. 3(c). Blue-shifts of several tens to one hundred meV in the plasmon resonance peaks are commonly observed following the loss of SERRS activity for $\sigma_{sca}(\omega)$ and $\sigma_{ext}(\omega)$. The blue-shifts occur simultaneously with the disappearance of SERRS activity [7,9,12]. As the origin of SERRS is the EM coupling between a plasmon and a molecular exciton [3,4], these simultaneous blue-shifts are considered to be the result of a loss of EM coupling energy by SERRS quenching. Indeed, we have quantitatively reproduced the blue-shifts in $\sigma_{sca}(\omega)$ as returning to the original plasmon resonance following the loss of EM coupling energy, including changes in the cross-sections of SERRS [12].

We consider the decrease in the EM coupling energy between the plasmon and molecular exciton resonance to be induced by an increase in the effective distance between the R6G molecules and the center of the hotspot having the highest mode density. Our previous analysis of a relationship between the intensity ratios of SERRS to surface enhanced fluorescence (SEF) and the degree of signal fluctuations demonstrated an increase in the effective distance of several angstroms based on the energy transfer from excited molecules to metal surfaces [6,33]. However, there is no direct evidence of such effective distance changes from e.g. tip-enhanced Raman scattering (TERS) measurements. Thus, we should consider other possibilities for the spectral changes in Fig. 3. The first possibility is a decrease in the number of R6G molecules in the hotspots in the case where $N$ is much larger than our estimation using the present coupling energy <450 meV as $N$ <5 [12]. The second possibility is anion-induced structural changes in dimers [34]. Another possibility is photo-induced melting of the dimers [35].

We now discuss the relationships among $\sigma_{sca}(\omega)$, $\sigma_{ext}(\omega)$, and $\sigma_{abs}(\omega)$ in the dimers. When $\sigma_{abs}(\omega) > \sigma_{sca}(\omega)$, as in Figs. 3(a1) and 3(a2), peaks appear in $\sigma_{abs}(\omega)$ near the peaks of $\sigma_{sca}(\omega)$ and $\sigma_{ext}(\omega)$. However, for $\sigma_{abs}(\omega) \sim \sigma_{sca}(\omega)$, as in Figs. 3(b1) and 3(b2), the appearance of such peaks in $\sigma_{abs}(\omega)$ becomes unclear. For $\sigma_{abs}(\omega) < \sigma_{sca}(\omega)$, as in Figs. 3(c1) and 3(c2), dips in $\sigma_{abs}(\omega)$ appear near the peaks of $\sigma_{sca}(\omega)$ and $\sigma_{ext}(\omega)$. These tendencies can be explained as the dimer-size-dependence of $\sigma_{sca}(\omega)$, $\sigma_{ext}(\omega)$, and $\sigma_{abs}(\omega)$ based on our discussion of the behavior of gold NPs in Fig. 2. The relationships $\sigma_{ext} \propto x$, $\sigma_{sca} \propto x^4$, and $\sigma_{abs}(\omega) = \sigma_{ext}(\omega) - \sigma_{sca}(\omega)$ predict a peak and a dip in $\sigma_{abs}(\omega)$ for a small and large dimer, respectively, because the scattering loss is more sensitive than the absorption loss in $\sigma_{ext}(\omega)$ as the dimer size increases [25]. Indeed, the peaks of $\sigma_{sca}(\omega)$ in Figs. 3(a1) and 3(a2) are much smaller than those in Fig. 3(c1) and 3(c2), reflecting the change in dimer sizes. The appearance of peaks and dips suggests that controlling the dimer size would enable the optical absorption of plasmon-exciton hybridized systems to be optimized.

We attempted to reproduce the changes in $\sigma_{sca}(\omega)$, $\sigma_{ext}(\omega)$, and $\sigma_{abs}(\omega)$ in Fig. 3 by decreasing the EM coupling energies. For this purpose, we used a modified coupled-oscillator model representing a strong coupling system between a plasmon of a silver NP dimer and an exciton of a dye molecule located at the dimer junction [12]. The model uses the Franck–Condon mechanism, which provides electron-vibration coupling, to yield a phonon replica of the exciton line [4]. Thus, the coupled-oscillator is composed of an oscillator representing a plasmon and multiple oscillators representing a molecular exciton and its phonon replicas. The equations of motion for the coupled

oscillators are:

$$\frac{\partial^2 x^p(t)}{\partial t^2} + \gamma^p \frac{\partial x^p(t)}{\partial t} + \omega^{p2} x^p(t) + \sum_{n=1}^{N} g_n \frac{\partial x_n^m(t)}{\partial t} = P^p(t), \qquad (1)$$

$$\frac{\partial^2 x_n^m(t)}{\partial t^2} + \gamma_n^m \frac{\partial x_n^m(t)}{\partial t} + \omega_n^{m2} x_n^m(t) - g_n \frac{\partial x^p(t)}{\partial t} = 0 \quad (n = 1, 2, 3, 4), \qquad (2)$$

where $x^p$ and $x_n^m$ are the coordinates of the plasmon and the $n$-th excitonic oscillation ($n = 1$ indicates an exciton and $n > 1$ indicates its phonon replicas), respectively; $\gamma^p$ and $\gamma_n^m$ are the line-widths of the plasmon and $n$-th excitonic resonance, respectively; $\omega^p$ and $\omega_n^m$ are the plasmon and $n$-th excitonic resonance frequencies, respectively; $g_n$ is the coupling rate between the plasmon and $n$-th excitonic resonance, and $P^p$ denotes the driving forces representing incident light [12]. We assume that the excitonic oscillators are entirely driven by the plasmon oscillator [36]. By assuming that $P^P(t) = P^P e^{-i\omega}$, $x^p(t)$ and $x_n^m(t)$ can be derived from Eqs. (1) and (2). The polarizability $\alpha = P^P x^P$ is then obtained as

$$\alpha \propto \frac{\prod_{n=1}^{N}\left(\omega_n^{m2} - \omega^2 - i\gamma_n^m \omega\right)}{(\omega^2 - \omega^{p2} + i\gamma^p \omega)\prod_{n=1}^{N}(\omega^2 - \omega_n^{m2} + i\gamma_n^m \omega) - \sum_{n=1}^{N}\omega^2 g_n^2 \dfrac{\prod_{k=1}^{N}(\omega^2 - \omega_k^{m2} + i\gamma_k^m \omega)}{(\omega^2 - \omega_n^{m2} + i\gamma_n^m \omega)}}. \qquad (3)$$

The coupling rate $g_n$ is determined by the oscillator strength of the electronic transition $f_n$ and the effective mode volume of hotspot $V$:

$$g_n = \left(\frac{1}{4\pi\varepsilon_r\varepsilon_0}\frac{\pi e^2 N f_n}{mV}\right)^{1/2}, \qquad (4)$$

where $\varepsilon_r$ (= 1.77) and $\varepsilon_0$ are the relative permittivities of the surrounding water and vacuum, respectively, $e$ is the elementary charge of an electron, and $m$ is the free electron mass [37]. We tentatively assume $N = 1$ based on our previous estimation [12]. The values of $f_n$ are determined from the absorption spectrum of R6G molecules [4,12]: $f_1 = 2m\omega_1^m d^2/(e^2\hbar)$, where $d$ (= 0.12 nm) is the dipole length of R6G [4], and $f_2 - f_4$ are obtained by multiplying $f_1$ by the ratios of the peak intensity of the Lorentzian curves between $f_1$ and $f_n$ [12].

We examine the observed spectral changes in $\sigma_{sca}(\omega)$, $\sigma_{ext}(\omega)$, and $\sigma_{abs}(\omega)$ using the calculated absorption, scattering, and extinction cross-sections $C_{sca}(\omega)$, $C_{ext}(\omega)$, and $C_{abs}(\omega)$ under the quasi-static approximation that the dimers are sufficiently small compared with the wavelength of the light. This approximation provides simple expressions for $C_{sca}(\omega)$ and $C_{ext}(\omega)$ [25]:

$$C_{sca}(\omega) \propto x^4 |\alpha|^2 \tag{5}$$

$$C_{ext}(\omega) \propto x \, \text{Im}(\alpha) \tag{6}$$

In the case of $x \ll 1$ (i.e., Rayleigh approximation), $C_{ext}(\omega) = C_{abs}(\omega)$ because $C_{abs}(\omega) \gg C_{sca}(\omega)$ regarding $C_{ext}(\omega) = C_{abs}(\omega) + C_{sca}(\omega)$. However, the scattering loss appearing in $\sigma_{ext}(\omega)$ in Fig. 3 indicates that the dimers are too large for the Rayleigh approximation to be applicable. Thus, we use $C_{abs}(\omega)$ under the 2nd order approximation to the scattering loss, which is given by

$$C_{\text{abs}}(\omega) \propto x \operatorname{Im}(\alpha) \left[ 1 - \frac{4x^3}{3} \operatorname{Im}(\alpha)^2 \right]. \tag{7}$$

The derivation of Eq. (7) is described in Ref. [25]. We evaluate the experimental spectral changes in $\sigma_{\text{sca}}(\omega)$, $\sigma_{\text{ext}}(\omega)$, and $\sigma_{\text{abs}}(\omega)$ during the loss of SERRS activity with $C_{\text{sca}}(\omega)$, $C_{\text{ext}}(\omega)$, and $C_{\text{abs}}(\omega)$ by decreasing $\hbar g_n$, because the origin of SERRS is the EM coupling between a plasmon and a molecular exciton [4,5]. Thus, the simultaneous blue-shifts with the loss of SERRS activity can be considered to be the result of losing the EM coupling energy. $\omega^p$ and $\gamma^p$ in Eq. (3) are taken from $\sigma_{\text{sca}}(\omega)$ after the loss of SERRS activity by assuming that $\hbar g_n = 0$. $\omega_n^m$ and $\gamma_n^m$ are taken from Ref. [12]. Note that Eqs. (5)-(7) are not applicable for large NPs because of the restriction of $x \ll 1$. Indeed, $\sigma_{\text{abs}}(\omega)$ derived by Eq. (7) with large $x$ have negative values, indicating the breakdown of the approximation in Eqs. (5)-(7) [25].

We can check the anti-crossing properties for $C_{\text{sca}}(\omega)$, $C_{\text{ext}}(\omega)$, and $C_{\text{abs}}(\omega)$ using Eqs. (5)–(7) with a coupling energy of $\hbar g_1 = 200$ meV. Figures 4(a) and 4(b) show the anti-crossing behavior of $C_{\text{sca}}(\omega)$ and $C_{\text{ext}}(\omega)$. Note that the spectral shapes in higher-energy regions (>2.4 eV) are complicated without showing the clear spectral splitting of the vacuum Rabi splitting in Fig. 1(b). The complexity is caused by multiple strong couplings between the plasmon and phonon replica of the molecular exciton generated by electron-vibration coupling, as described by Eq. (3) [12]. The reason for the spectral changes, and even the detuning between plasmon resonance and molecular resonance, is the overlapping of both the resonance spectra owing to their broad spectral widths. The reason for $C_{ext}(\omega)$ exhibiting broader spectral lines than $C_{sca}(\omega)$ is that $C_{ext}(\omega) \propto \text{Im}(\alpha)$ and $C_{sca}(\omega) \propto |\alpha|^2$, as indicated in Eqs. (5) and (6). Additionally, the intensity of $C_{ext}(\omega)$ is less sensitive to spectral peak shifts than that of $C_{sca}(\omega)$ because $C_{ext}(\omega) \propto \omega$ and $C_{sca}(\omega) \propto \omega^4$, as indicated in Eqs. (5) and (6).

The anti-crossing properties of $C_{abs}(\omega)$ exhibit some dimer-size dependence owing to the scattering loss $(4x^3/3)\text{Im}(\alpha)^2$, as described in Eq. (7). Thus, we examined the anti-crossing properties of $C_{abs}(\omega)$ by changing the size parameters. Figures 4(c) and 4(d) show the anti-crossing behavior of $C_{abs}(\omega)$ calculated with $x = 0.29$ and $x = 0.37$, respectively. Note that $C_{abs}(\omega)$ in Fig. 4(c) shows similar anti-crossing behavior as $C_{ext}(\omega)$ in Fig. 4(b) under the rather negligible effect of scattering loss. However, Fig. 4(d) exhibits dips rather than peaks in the hybridized spectral lines. The dips are the result of large scattering loss in Eq. (7), indicating that scattering loss is no longer negligible for large dimers.

We verify this qualitative discussion in Fig. 5 by comparing the experimental and calculation results under the quasi-static approximation. Figures 5(a)–5(c) show the experimental changes in $\sigma_{sca}(\omega)$, $\sigma_{ext}(\omega)$, and $\sigma_{abs}(\omega)$ during SERRS quenching for the three dimers in Figs. 3(a)–3(c). Figures 5(d)–5(f) show the calculated changes in $C_{sca}(\omega)$, $C_{ext}(\omega)$, and $C_{abs}(\omega)$ given by Eqs. (5)–(7), respectively, while decreasing $\hbar g_n$ in Eq. (3) to reproduce the experimental changes. The blue-shifts and increases in the peaks of $\sigma_{sca}(\omega)$ and $\sigma_{ext}(\omega)$ following the loss of SERRS activity are well reproduced by $C_{sca}(\omega)$ and $C_{ext}(\omega)$ with decreasing $\hbar g_n$, indicating that the spectral changes can be explained as a return to the original plasmon resonance from hybridized resonance following the loss of EM coupling between plasmon resonance and molecular exciton resonance. The values of $\hbar g_1$ reproducing the experimental changes of <500 meV may be acceptable for determining the maximum value of $\hbar g_1$ [11,12]. Equation (4) indicates that 500 meV corresponds to $V = (0.81)^3 N$ nm$^3$ at 2.15 eV. This value of $V$ indicates that a sub-nanometer cavity is realized at SERRS hotspots by assuming $N \sim 1$. Such an extremely small plasmonic cavity has also been deduced from rigorous analysis of the spectral changes in plasmon resonance using plasmonic NPs on plasmonic substrates [38]. Furthermore, interesting phenomena that cannot be explained by excluding such small plasmonic cavities have been reported, e.g., TERS imaging of the internal structures of SMs [39]. We have also successfully reproduced SERRS spectra using this value of $\hbar g_1$ [12]. The changes in intensity of the peaks in $\sigma_{sca}(\omega)$ peaks are greater than those for $\sigma_{\text{ext}}(\omega)$. This tendency can also be observed in the changes in $C_{\text{sca}}(\omega)$ and $C_{\text{ext}}(\omega)$, indicating that the changes in the degree of intensity with blue-shifts are explained by $\sigma_{\text{ext}} \propto x$ and $\sigma_{\text{sca}} \propto x^4$ in Eqs. (5) and (6), because $x$ is a function of $\omega$, i.e., $x = \dfrac{\omega n_{\text{ref}} d}{2c}$.

The dimer-by-dimer variations in spectral changes in $\sigma_{\text{abs}}(\omega)$ look more complicated than those in $\sigma_{\text{sca}}(\omega)$ and $\sigma_{\text{ext}}(\omega)$. For example, Fig. 5(a3) shows a blue-shift in a peak of $\sigma_{\text{abs}}(\omega)$ without any increase in intensity, Fig. 5(b3) shows a blue-shift in a peak of $\sigma_{\text{abs}}(\omega)$ with a decrease in intensity, and Fig. 5(c3) shows a blue-shift in a dip (not a peak) of $\sigma_{\text{abs}}(\omega)$ with a decrease in intensity. We examine the complicated changes in $\sigma_{\text{abs}}(\omega)$ with respect to dimer size using Eq. (7). From the values of $\sigma_{\text{sca}}(\omega)$ in Fig. 3, we can determine that the dimer size in Fig. 5(a) is less than that in Fig. 5(b), which is in turn less than that in Fig. 5(c). Thus, we set the value of $x$ in Eq. (7) for $C_{\text{abs}}(\omega)$ in Fig. 5(d3) to be less than that in Fig. 5(e3), which is in turn set to be less than that in Fig. 5(f3); namely, we use values of 0.29, 0.35, and 0.42, corresponding to diameters of 46, 54, and, 64 nm under the assumption of approximately spherical shapes. Note that the derivation of dimer sizes is just an order estimation, showing that our approximation is not so far from the real situation. The spectra of $C_{\text{abs}}(\omega)$ in Figs. 5(d3)–5(f3) effectively reproduce the complicated changes in $\sigma_{\text{abs}}(\omega)$. The calculation demonstrates that the peak in $C_{\text{abs}}(\omega)$ changes to a dip as the size parameter increases, indicating that the dips in $\sigma_{\text{abs}}(\omega)$ appear due to scattering loss, which is more sensitive to dimer size than extinction loss, as indicated in Eq. (7). We now comprehensively explain the spectral changes in $\sigma_{\text{sca}}(\omega)$, $\sigma_{\text{ext}}(\omega)$, and $\sigma_{\text{abs}}(\omega)$ by SERRS quenching using $C_{\text{sca}}(\omega)$, $C_{\text{ext}}(\omega)$, and $C_{\text{abs}}(\omega)$ with decreasing $\hbar g_n$. In particular, the variations in spectral changes in $\sigma_{\text{abs}}(\omega)$ can be explained as the size-parameter dependence of $C_{\text{abs}}(\omega)$.

We applied the above evaluation of $\sigma_{\text{sca}}(\omega)$, $\sigma_{\text{ext}}(\omega)$, and $\sigma_{\text{abs}}(\omega)$ using $C_{\text{sca}}(\omega)$, $C_{\text{ext}}(\omega)$, and $C_{\text{abs}}(\omega)$ to SERRS active 14 dimers to check the validity of our explanation. Figures 6(a) and 6(b) show the changes in the peak positions of $\sigma_{\text{sca}}(\omega)$ and $\sigma_{\text{ext}}(\omega)$ during the loss of SERRS activity; their peak cross-sections are normalized by their values after the loss of SERRS activity. Figure 6(c) shows changes in the peak or dip positions of $\sigma_{\text{abs}}(\omega)$ during the loss of SERRS activity; their peak or dip cross-sections are normalized by their values before the loss of SERRS activity. Note that the increases in the ratios for $\sigma_{\text{sca}}(\omega)$ are generally larger than those for $\sigma_{\text{ext}}(\omega)$, and the decreases in the ratios for $\sigma_{\text{abs}}(\omega)$ showing dips are generally larger than for those showing peaks. We examine these tendencies using $C_{\text{sca}}(\omega)$, $C_{\text{ext}}(\omega)$, and $C_{\text{abs}}(\omega)$. Figures 6(d) and 6(e) show the changes in the peak positions of $C_{\text{sca}}(\omega)$ and $C_{\text{ext}}(\omega)$ as $\hbar g_1$ decreases from 500 meV to 0 meV, with the peak intensities normalized by their values at $\hbar g_1 = 0$ meV. Figure 6(f) shows the changes in the peak or dip positions of $C_{abs}(\omega)$ as the coupling energy decreases from 500 meV to 0 meV, with the peak or dip intensities normalized by their values at $\hbar g_1 = 500$ meV. We use values of $x = 0.29$ and 0.37 to generate peaks and dips in $C_{abs}(\omega)$, respectively. Both $\sigma_{sca}(\omega)$ and $\sigma_{ext}(\omega)$ exhibit blue-shifts as the intensity increases. The changes in the degree of intensity for $\sigma_{sca}(\omega)$ and $\sigma_{ext}(\omega)$ are well reproduced by $C_{sca}(\omega)$ and $C_{ext}(\omega)$, indicating that the difference is due to $\sigma_{ext} \propto x$ and $\sigma_{sca} \propto x^4$. The SERRS active dimers with peaks in $\sigma_{sca}(\omega)$ and $\sigma_{ext}(\omega)$ in higher-energy regions (>1.9 eV) exhibit larger blue-shifts and ratio increases for $\sigma_{sca}(\omega)$ and $\sigma_{ext}(\omega)$ than the dimers having peaks in lower-energy regions (<1.86 eV). $C_{sca}(\omega)$ and $C_{ext}(\omega)$ exhibit similar tendencies, indicating that this is caused by detuning between the plasmon resonance and molecular exciton resonance. The SERRS active dimers with peaks in $\sigma_{abs}(\omega)$ exhibit much smaller ratio decreases than the dimers showing dips in $\sigma_{abs}(\omega)$. $C_{abs}(\omega)$ also reproduces this property, indicating that the difference in the scattering loss $\propto x^4$ can generate the properties required for $C_{abs}(\omega) = C_{ext}(\omega) - C_{sca}(\omega)$. This consistency between the experiments and calculations indicates that the observed complicated dimer-by-dimer variations in spectral changes are reproduced using just three parameters: the original peak energies of plasmon resonance, coupling energies, and dimer sizes under a quasi-static approximation.

## IV. SUMMARY

In this work, we investigated spectral changes in $\sigma_{sca}(\omega)$, $\sigma_{ext}(\omega)$, and $\sigma_{abs}(\omega)$ of a single strong coupling system composed of a silver NP dimer and a few dye molecules during the quenching of SERRS. The $\sigma_{abs}(\omega)$ were successfully obtained by subtracting $\sigma_{sca}(\omega)$ from $\sigma_{ext}(\omega)$, and the spectral changes in $\sigma_{sca}(\omega)$, $\sigma_{ext}(\omega)$, and $\sigma_{abs}(\omega)$ were correctly evaluated using a classical hybridization model composed of a plasmon and a molecular exciton including a molecular multi-level property. The blue-shifts in peak energies and increasing peak cross-sections in $\sigma_{sca}(\omega)$ and $\sigma_{ext}(\omega)$, respectively, during SERRS quenching were effectively reproduced by decreasing the coupling energy in the model. The peak positions in $\sigma_{sca}(\omega)$ and $\sigma_{ext}(\omega)$ appear as peaks or dips in $\sigma_{abs}(\omega)$ depending on the degree of scattering loss, which reflects the dimer sizes. This property was also reproduced well using the size parameters in the model.

Finally, we discuss how the current result contributes to studies of photophysical and photochemical phenomena in strong coupling systems. As explained in the introduction, the molecular resonance energy can be reduced by EM coupling, because the reduction of $\frac{\hbar g_1}{\sqrt{N}}$ for each molecule is non-negligible under the condition $N \sim 1$ [15]. Thus, a conventional non-resonant excitation condition may become resonant following this reduction, resulting in, e.g., nonlinear optical responses such as vibrational pumping and Rabi splitting driven by excitation photon which energy is energy below the conventional threshold [13,16,17]. However, we should check the absorption spectra of strong coupling systems to determine whether hybridized resonance peaks in the scattering or extinction spectra are close to the excitation photon energies, because the absorption cross-section may attain a minimum at the peak positions in the scattering or extinction spectra, as in Fig. 5(c) e.g. Ref. 40. This situation results in inefficient photochemical and photophysical phenomena. We plan to apply absorption spectroscopy to unique phenomena such as ultrafast surface-enhanced fluorescence [41] and unique plasmonic systems such as one-dimensional SERRS hotspots [42] to further explore the effect of the decrease in resonance energy caused by strong coupling.

## ACKNOWLEDGMENT

This work was supported by JSPS KAKENHI Grant-in-Aid for Scientific Research (C) Number 18K04988.

**Figure captions**

FIG. 1 (a) (Color online) Strong coupling between plasmon resonance and molecular resonance. |g> and |e> are the ground and excited states of a two-level system, respectively. |0> and |1> are the zero-photon and one-photon state of a plasmonic resonator. $2\hbar g$ indicates the energy split of the excited state of a strong coupling system between the two-level system and the plasmonic resonator. Red and blue arrows indicate the two resonance energies of the strong coupling system. (b) Plasmon resonance spectral calculations for the examination of spectral splitting. Calculated spectra using $\hbar g$ from 500 meV to 0 meV at intervals of 100 meV. Red and blue arrows indicate the two resonance energies of the strong coupling system in (a). (c)–(e) Dark-field, bright-field, and SERRS images of silver NPs and their aggregates dispersed on a glass plate, respectively. Scale bars are 5 μm. (f) with (g) and (h) with (i) Plasmon resonance light scattering spectra (blue lines) and SERRS spectra (red lines) of two silver NP dimers measured by SEM. Scale bars are 100 nm.

FIG. 2 (a)–(c) Calculated spectra of $\sigma_{ext}(\omega)$ (red lines), $\sigma_{sca}(\omega)$ (blue lines), and $\sigma_{abs}(\omega)$ (green lines) of gold NPs with diameters of 30 nm (a), 70 nm (b), and 100 nm (c). (d) Experimental relationships between peak energies and cross-sections at those energies for $\sigma_{\text{ext}}(\omega)$ (red marks), $\sigma_{\text{sca}}(\omega)$ (blue marks), and $\sigma_{\text{abs}}(\omega)$ (green marks) for gold NPs with average diameters of 60 nm (open circles), 80 nm (open diamonds), and 100 nm (open triangles). Calculated relationships between peak energies and cross-sections at those energies for $\sigma_{\text{ext}}(\omega)$ (red closed circles with line), $\sigma_{\text{sca}}(\omega)$ (blue closed circles with line), and $\sigma_{\text{abs}}(\omega)$s (green closed circles with line) for gold NPs of diameters 20–100 nm. (e)–(g) Experimental and calculated $\sigma_{\text{ext}}(\omega)$ (red lines), $\sigma_{\text{sca}}(\omega)$ (blue lines), and $\sigma_{\text{abs}}(\omega)$ (green lines) for gold NPs with average diameters of 60 nm (e), 80 nm (f), and 100 nm (g). Experimental and calculated spectra are indicated by pale and dark lines, respectively.

FIG. 3 (a)–(c) Spectra of $\sigma_{\text{ext}}(\omega)$ (red lines), $\sigma_{\text{sca}}(\omega)$ (blue lines), and $\sigma_{\text{abs}}(\omega)$ (green lines) of three silver NP dimers. White bar indicates loss by laser notch filter. (a1)–(c1) Spectra of cross-sections for the three dimers showing SERRS activity. (a2)–(c2) Spectra of cross-sections for the three dimers after losing SERRS activity. Dotted lines in (a1)–(c1) and (a2)–(c2) indicate blue-shifts of spectra following loss of SERRS activity. (a3)–(c3) SERRS spectra before (red lines) and after (black lines) loss of SERRS activity for the three dimers. Insets are enlarged SERRS spectra showing Raman lines of R6G molecules.

FIG. 4 Anti-crossing properties appearing in hybridized resonance spectra calculated by Eqs. (5)–(7) under the conditions with $\hbar g_1 = 200$ meV for (a) $C_{\text{sca}}(\omega)$, (b) $C_{\text{ext}}(\omega)$, (c) $C_{\text{abs}}(\omega)$ with $x = 0.29$, and (d) $C_{\text{abs}}(\omega)$ with $x = 0.37$. $\hbar\gamma^p = 300$ meV. The values of $\omega_n^m$ and $\gamma_n^m$ are taken from Ref. [12].

FIG. 5 (a)–(c) Experimental spectra of $\sigma_{\text{sca}}(\omega)$ (a1–c1, blue red with black lines), $\sigma_{\text{ext}}(\omega)$ (a2–c2, red blue with black lines), and $\sigma_{\text{abs}}(\omega)$ (a3–c3, green with black lines) of three silver NP dimers before (red, blue, and green lines) and after (black lines) loss of SERRS activity. White bar indicates loss by laser notch filter. (d)–(f) Calculated spectra of $C_{\text{sca}}(\omega)$ (d1–f1, blue with black lines), $C_{\text{ext}}(\omega)$ (d2–f2, red with black lines), and $C_{\text{abs}}(\omega)$ (d3–f3, green with black lines) under the conditions that $\hbar\omega^p = 2.2$ eV (d), 2.1 eV (e), and 2.0 eV (f) with $\hbar\gamma^p = 250$ meV (d, e) and 300 meV (f) in Eq. (3) for reproducing experimental spectral changes in (a)–(c) by reducing coupling energies from $\hbar g_1 = 250$ meV (d), 400 meV (e), and 350 meV (f) to 0. The values of $x$ in Eq. (7) for (d3)–(f3) are 0.29, 0.35, and 0.42, respectively. The arrows in (c3) and (f3) indicate the positions of spectral dips.

FIG. 6 (a)–(c) Experimental changes in peak (or dip) energies and normalized cross-sections at those energies during SERRS quenching for $\sigma_{sca}(\omega)$ (a), $\sigma_{ext}(\omega)$ (b), and $\sigma_{abs}(\omega)$ (c) of 14 silver NP dimers before (blue, red, and green open circles or squares) and after (black open circles or squares) loss of SERRS activity. $\sigma_{sca}(\omega)$ and $\sigma_{ext}(\omega)$ are normalized by the values before SERRS quenching. $\sigma_{abs}(\omega)$ is normalized by the values after SERRS quenching. The directions of changes are indicated by black dashed arrows. Note that changes in dip energies and $\sigma_{abs}(\omega)$ are indicated by open squares in (c). (d)–(f) Calculated changes in peak energies and normalized intensities at those energies in $C_{sca}(\omega)$ (d), $C_{ext}(\omega)$ (e), and $C_{abs}(\omega)$ (f) with blue, red, and green open circles or squares, respectively, using Eqs. (5)–(7) by reducing $\hbar g_1$ from 500 meV to 0 meV under the conditions that $\hbar \omega^p$ = 2.3–1.8 eV with $\hbar \gamma^p$ = 250 meV in Eq. (3). The directions of changes are indicated by black dashed arrows. In Eq. (7) for (f), $x$ takes values of 0.29 and 0.37. Note that changes in dip energies and $C_{abs}(\omega)$ are indicated by squares in (f). $C_{sca}(\omega)$ and $C_{ext}(\omega)$ are normalized by the values at $\hbar g_1$ = 0 meV. $C_{abs}(\omega)$ is normalized by the values at $\hbar g_1$ = 500 meV.

Fig. 1

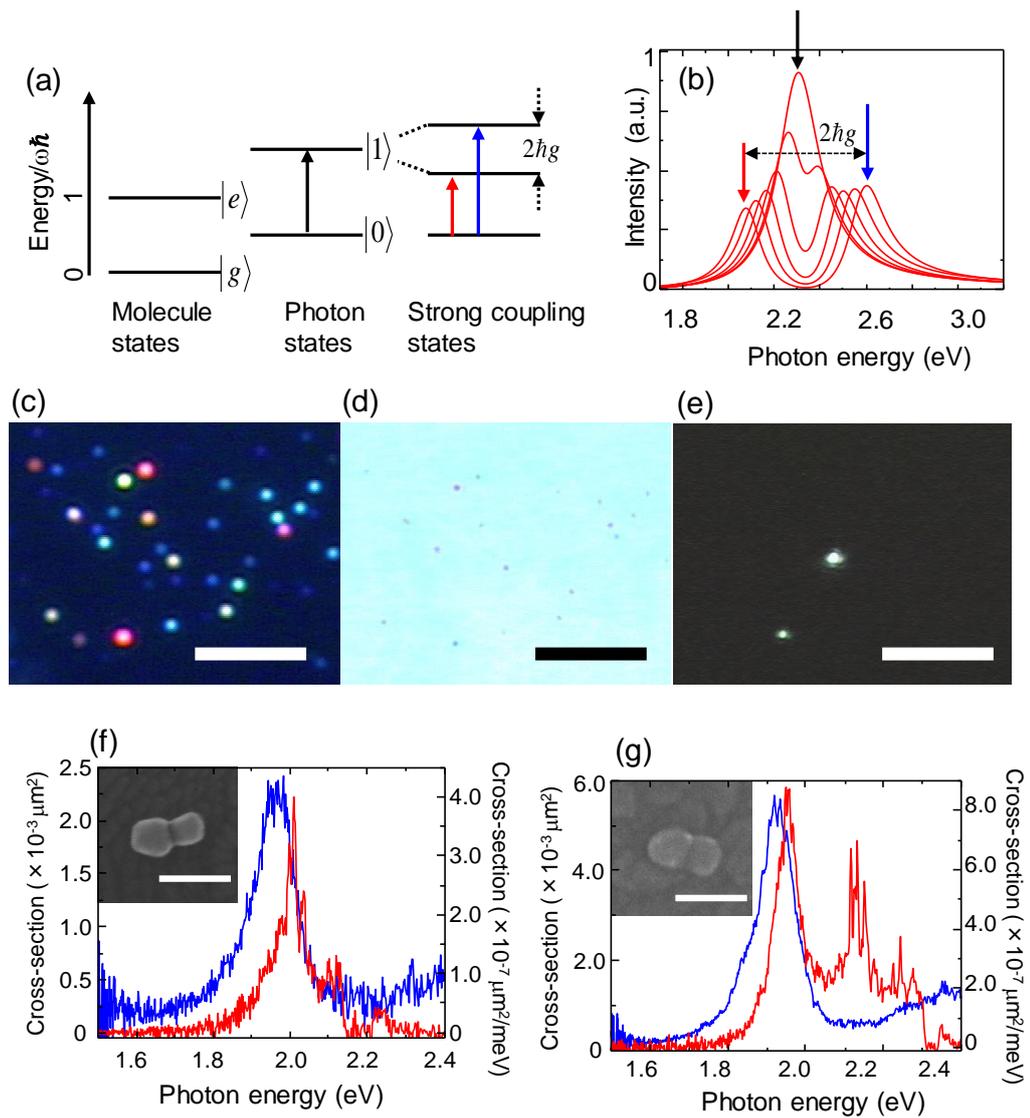

Fig. 2

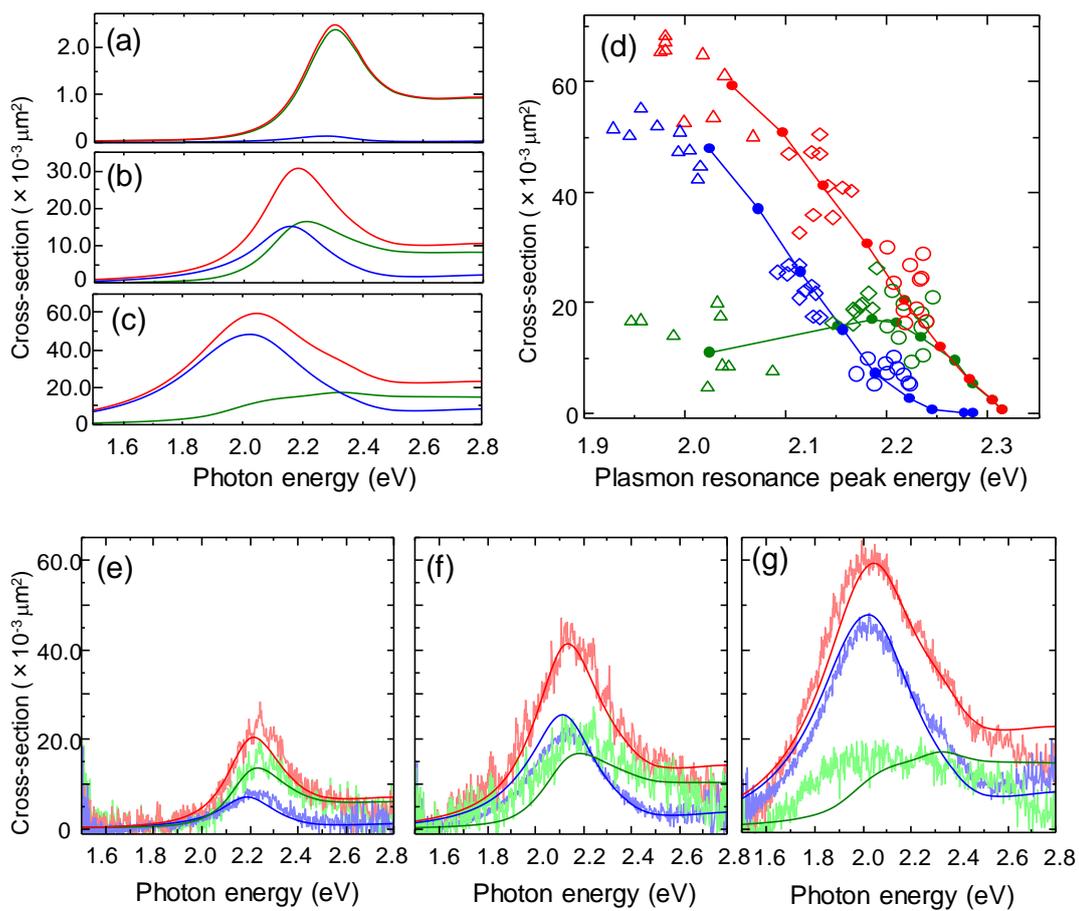

Fig. 3

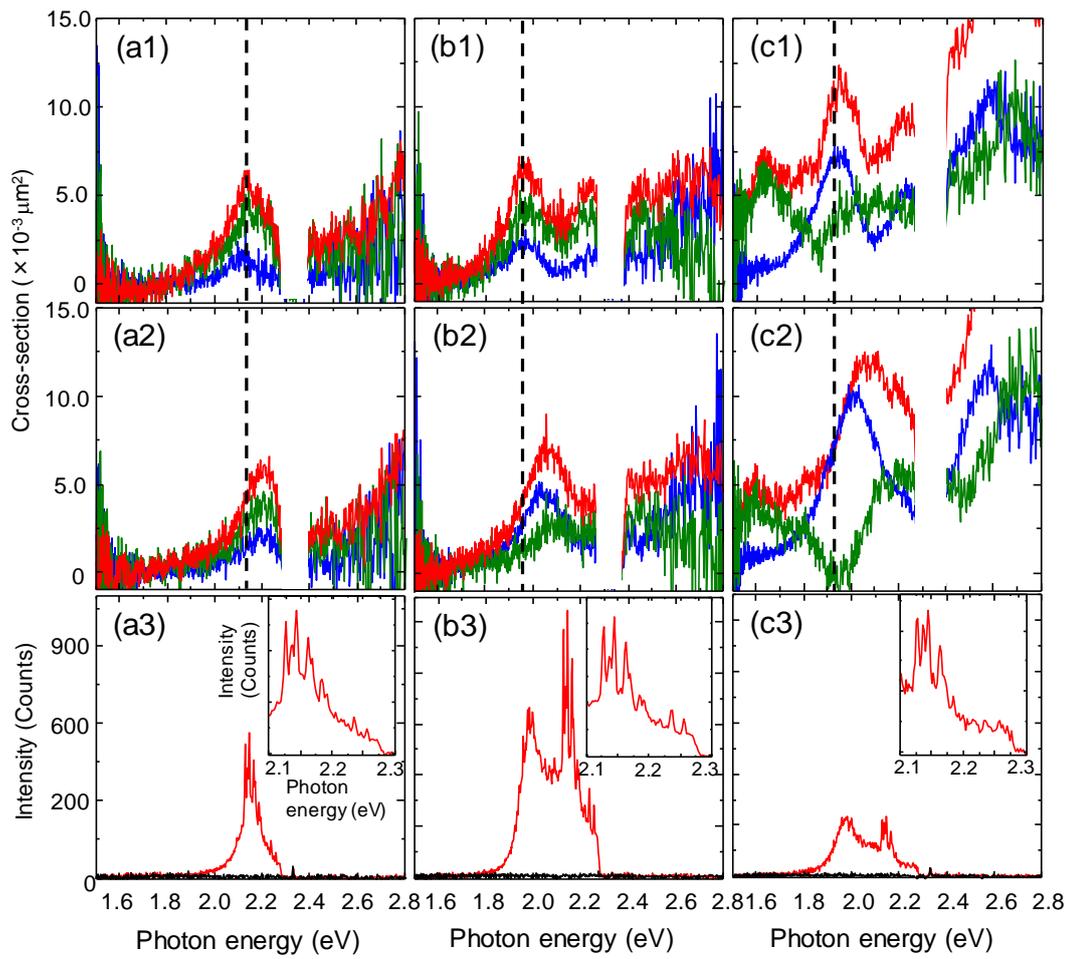

Fig. 4

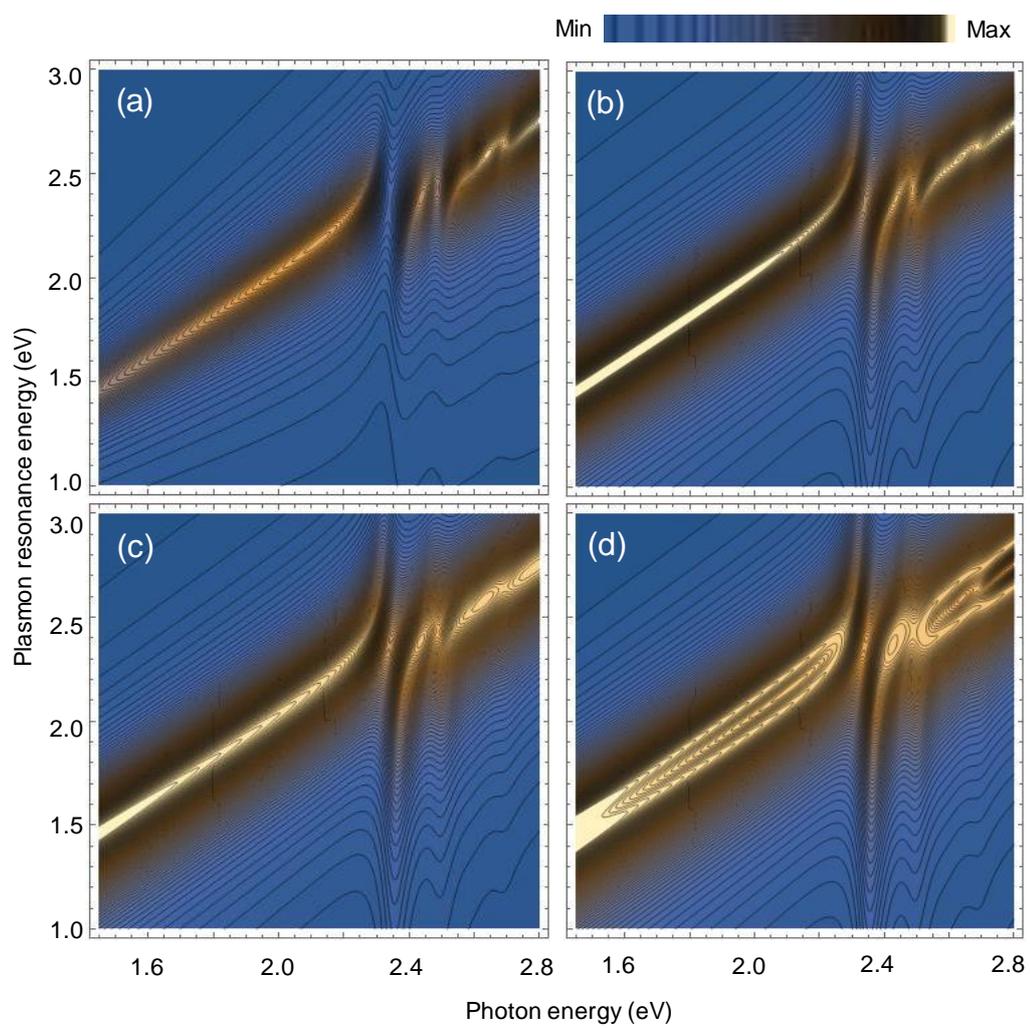

Fig. 5

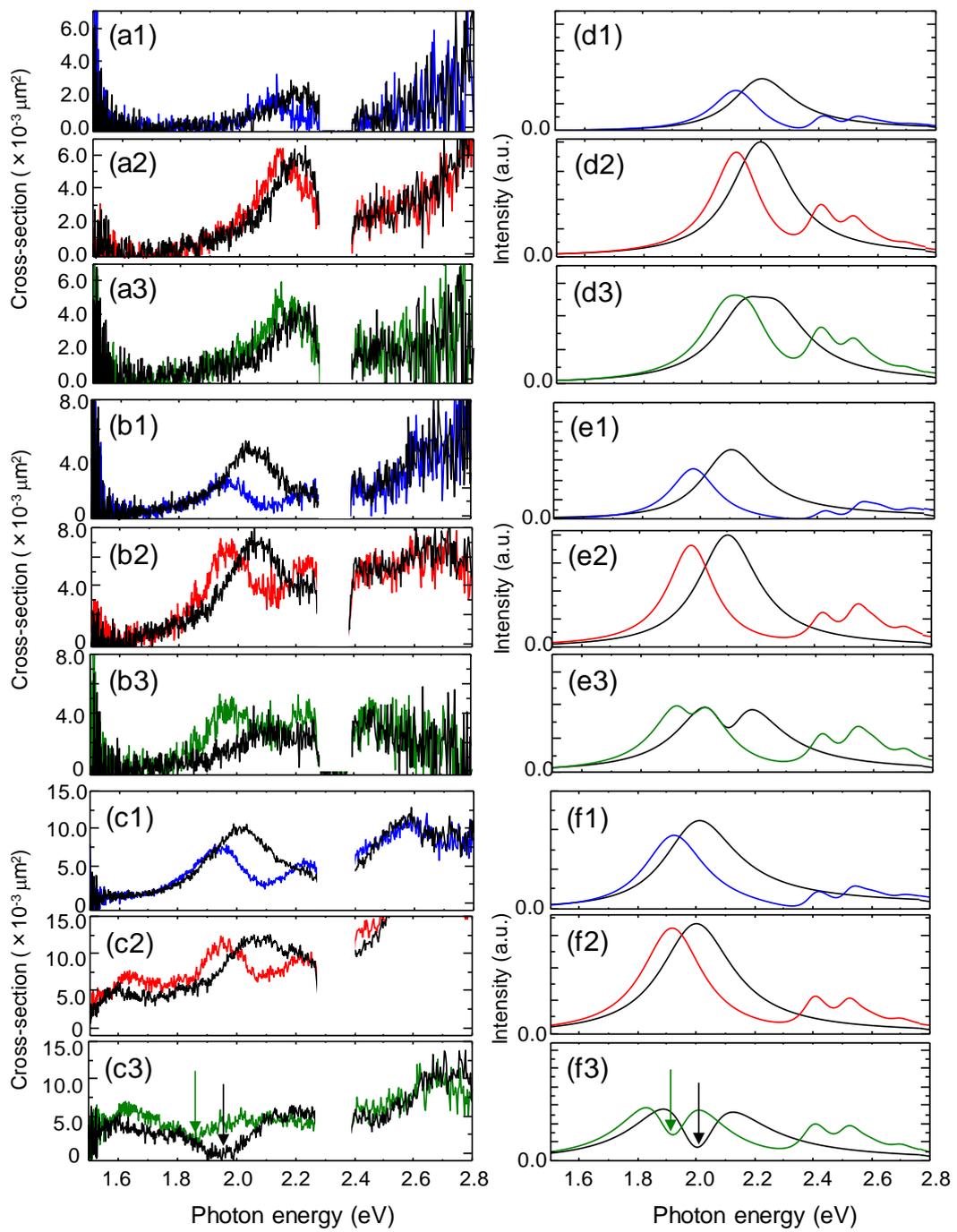

Fig. 6

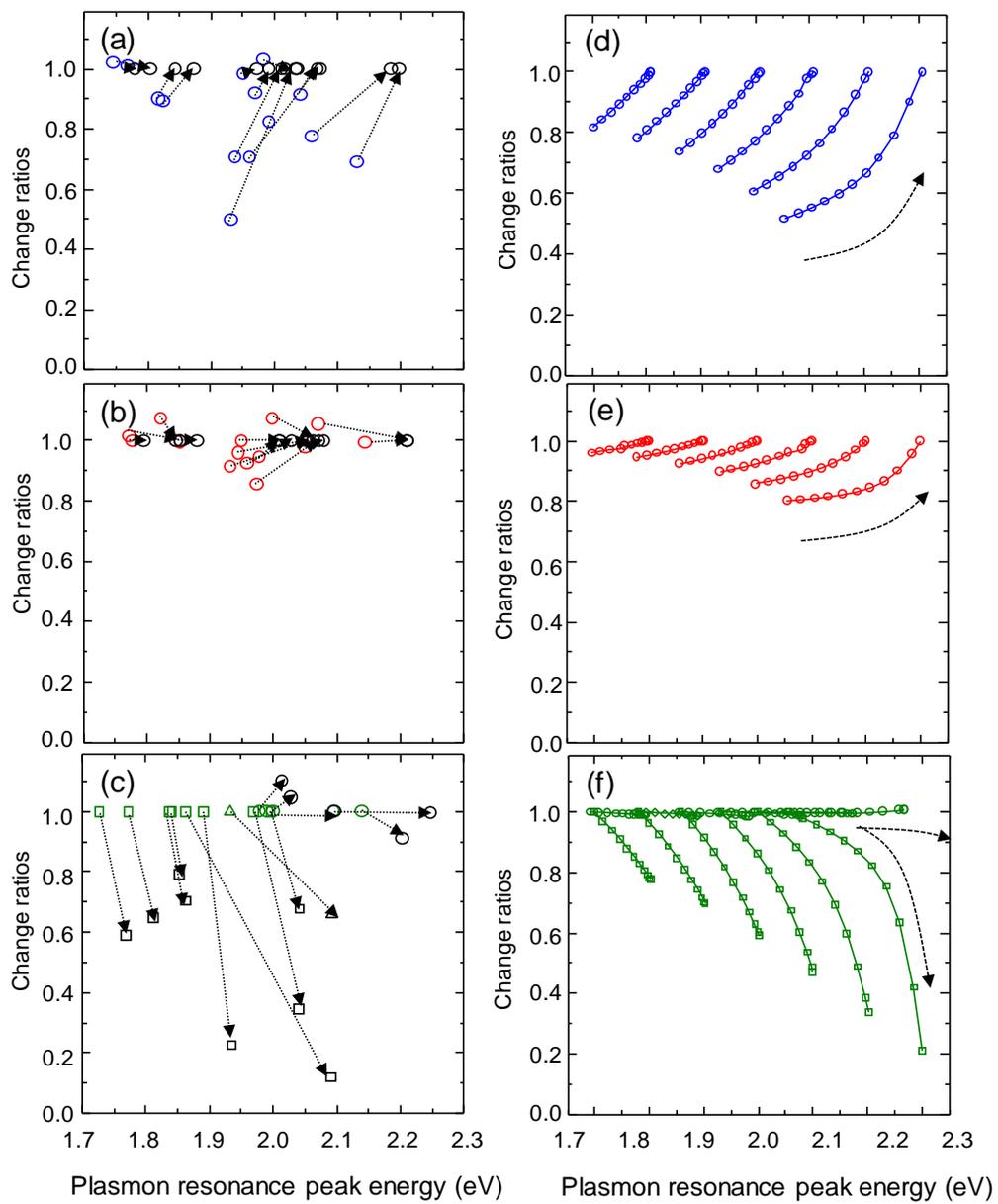